\documentclass[12pt,prl,superscriptaddress,showpacs,amsmath,showkeys,preprint]{revtex4}
\usepackage{graphicx}

\begin{document}

\title{Anderson Localization in Disordered Vibrating Rods}

\author{J.~Flores}
\affiliation{Instituto de F\'{\i}sica, Universidad Nacional Aut\'onoma de M\'exico, P.O. Box 20-364, 01000 M\'exico, D. F., Mexico}

\author{L.~Guti\'errez}
\affiliation{Instituto de Ciencias F\'isicas, Universidad Nacional
Aut\'onoma de M\'exico, P.O. Box 48-3, 62251 Cuernavaca, Mor., Mexico}

\author{R.~A.~\surname{M\'endez-S\'anchez}}
\affiliation{Instituto de Ciencias F\'isicas, Universidad Nacional
Aut\'onoma de M\'exico, P.O. Box 48-3, 62251 Cuernavaca, Mor., Mexico}

\author{G.~Monsivais}
\affiliation{Instituto de F\'{\i}sica, Universidad Nacional Aut\'onoma de M\'exico, P.O. Box 20-364, 01000 M\'exico, D. F., Mexico}

\author{P.~Mora}
\affiliation{Instituto de F\'{\i}sica, Universidad Nacional Aut\'onoma de M\'exico, P.O. Box 20-364, 01000 M\'exico, D. F., Mexico}

\author{A.~Morales}
\affiliation{Instituto de Ciencias F\'isicas, Universidad Nacional
Aut\'onoma de M\'exico, P.O. Box 48-3, 62251 Cuernavaca, Mor., Mexico}

\begin{abstract}
We study, both experimentally and numerically, the Anderson localization phenomenon in torsional waves of a disordered elastic rod, which consists of a cylinder with randomly spaced notches. We find that the normal-mode wave amplitudes are exponentially localized as occurs in disordered solids. The localization length is measured using these wave amplitudes and it is shown to decrease as a function of frequency. The normal-mode spectrum is also measured as well as computed, so its level statistics can be analyzed. Fitting the nearest-neighbor spacing distribution a level repulsion parameter is defined that also varies with frequency. The localization length can then be expressed as a function of the repulsion parameter. There exists a range in which the localization length is a linear function of the repulsion parameter, which is consistent with Random Matrix Theory. However, at low values of the repulsion parameter the linear dependence does not hold.
\end{abstract}

%

\maketitle

The Anderson localization phenomenon is a very important subject in condensed matter physics since it is crucial to understand the transport properties of materials. As a matter of fact, the original work of Anderson~\cite{Anderson} is among the most cited papers in twentieth century physics and it is at the core of many papers, not only in solid state studies but also in optics, cold atomic gases, microwaves and acoustics~\cite{BertolottiGottardoWiersmaGhulinyanPavesi,StorzerGrossAegerterMaret,TopolancikIlicVollmer,SchwartzBartalFishmanSegev,HuStrybulevychPageSkipetrovVanTiggelen,Lahinietal,Lietal,LagendijkVanTiggelenWiersma,FaezStrybulevychPageLagendijkVanTiggelen,BodyfeltZhengKottosKuhlStockmann,LemarieLignierDelandeSzriftgiserGarreau}. The theory of Anderson localization studies the alterations brought about on the localization of the electronic wave functions by disorder in the system. In a perfect lattice a band spectrum arises with extended wave functions for the allowed energy levels. However,  if the system presents random imperfections, for example the presence of strange atoms in an otherwise perfect structure or when there are unit cells of different size, wave functions can be localized, affecting the transport properties of the system. 

\begin{figure}
\includegraphics[width=\columnwidth]{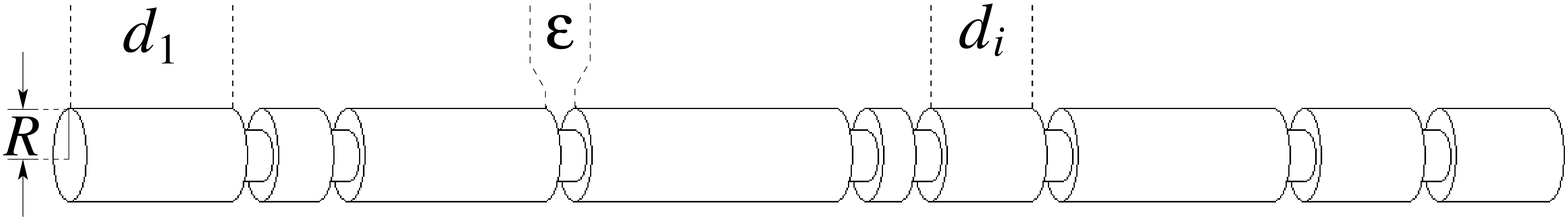}
\caption{One-dimensional rod used to measure localization. In the experiment, the number of rods $N=50$, $R=1.28$~cm, $d=\langle d_i \rangle=7.2$~cm, and $\epsilon=1.016$~mm; the values $\Delta=0.35$, $\eta=0.65$ were used. The largest frequency considered is less than $100$~kHz so the lowest wavelength $\lambda_\mathrm{min}=\frac{c}{f_\mathrm{max}} > 2 R$ and the system behaves indeed as 1D. The value $c=3140$~$\frac{m}{s}$ was measured for torsional waves in the aluminum alloy we have used.}
\label{Fig.System}
\end{figure}

Anderson localization can also be studied using elastic vibrating systems, such as the quasi one-dimensional rod shown in Fig.~\ref{Fig.System}. The system consists of $N$ rods of radius $R$ with lengths $d_i$, $i=1,\dots,N$, joined by smaller cylinders of length $\epsilon\ll d_i$, $\forall i$, and radius $r=\eta R$, where the coupling constant $\eta$ is such that $0<\eta<1$. According to the nature of the family of numbers $\{d_i\}$ different phenomena are observed. We should remark that we measure not only the normal-mode frequencies but also the wave amplitudes. In this sense, the analysis of elastic vibrations is more complete than what can be done in the quantum-mechanical or optical cases, since for such systems wave functions cannot, in general,  be observed. 

In this Letter we shall study torsional vibrations in a disordered rod by taking the family $\{d_i\}$ as a set of uncorrelated random numbers with a uniform distribution in the interval $[d(1-\Delta), d(1+\Delta)]$, where $d=\langle d_i\rangle$ is the average of $d_i$ and $\Delta$ measures the disorder. This allows us to study the Anderson localization phenomenon in elastic systems; in particular, we can observe how the wave amplitudes decay exponentially with a localization length $\xi$.

To perform the measurements we used the electromagnetic acoustic transducer (EMAT) developed by us~\cite{Moralesetal}. The EMAT consists of a permanent magnet and a coil, and can be used either to detect or excite the oscillations. The transducer operates through the interaction of eddy currents in the metallic rod with a permanent magnetic field. According to the relative position of the magnet and the coil, the EMAT can either excite or detect selectively compressional, torsional or flexural vibrations. Used as a detector, the EMAT measures acceleration. The experimental setup is described in detail in Ref.~\cite{Gutierrezetal}. This transducer has the advantage of operating without mechanical contact with the rod. This is crucial to avoid perturbing the shape of the localized wave amplitudes.

\begin{figure}
\includegraphics[width=\columnwidth]{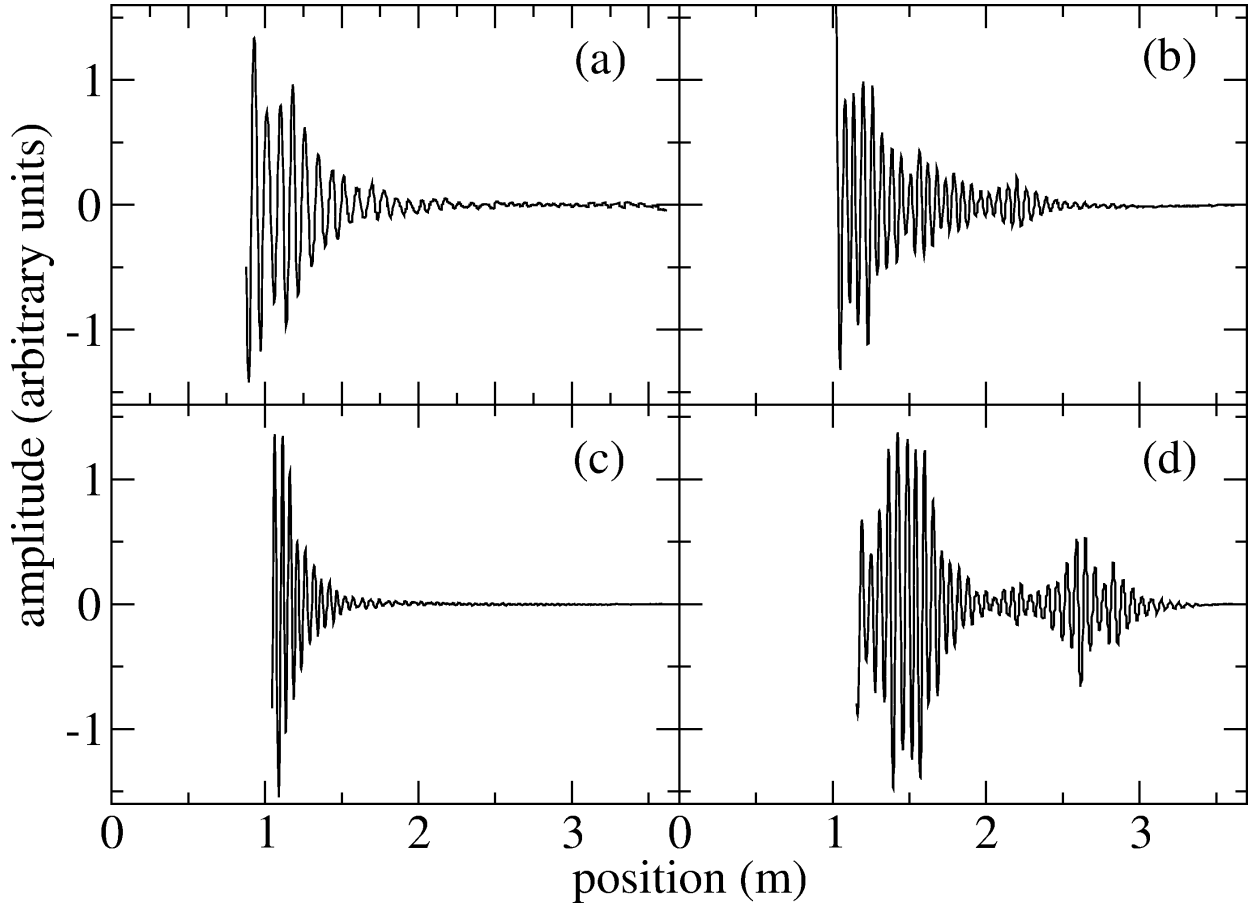}
\caption{Examples of experimental wave amplitudes for the disordered rod of Fig.~\ref{Fig.System}. The localization length $\xi$ is extracted from the envelopes of these plots of the amplitudes as a function of the position.}
\label{Fig.WaveAmplitudes}
\end{figure}

Before comparing the experimental data with the theoretical calculations, we shall consider what we will call an independent rod model~\cite{Gutierrezetal}: The small rods of length $d_i$ are independent from one another when $\eta \rightarrow 0$. In this case, the $i$-th rod is excited when the driving force has a frequency $f$ equal to $f_n^{(i)}=\frac{n c}{2 d_{i}}$, where $c$ is the speed of torsional waves and $n$ is an integer number. The other rods are in general not excited since $d_j$ for $j\ne i$ is usually different from $d_i$. The amplitude of the vibration decreases when one moves away from the $i$-th rod so the wave is localized. It could also happen that some other length, say $d_j$, can be almost equal to $d_i$. In this case the amplitude could then present two maxima. The first case is shown, for example, in Figs.~\ref{Fig.WaveAmplitudes}(a) and~\ref{Fig.WaveAmplitudes}(c), while in Figs.~\ref{Fig.WaveAmplitudes}(b) and~\ref{Fig.WaveAmplitudes}(d) the second case is apparent. The independent rod model therefore provides a qualitative argument to understand why all normal modes of a disordered rod are localized. Furthermore, when the disorder is very small, that is, when $\Delta \ll 1$, the same argument shows that almost all the rods can be excited with a driving force of frequency $f\sim \frac{c}{2d}$ and the localization length $\xi$ grows and could even exceed the total length of the complete rod; the amplitudes are then extended. 

If the cylinders have random lengths the elastic vibrations are waves on a random structure, which is analogous to what happens with the Schr\"odinger wave functions in a random potential, frequency playing the role of energy. The independent rod model then shows that introducing disorder in $\{d_i\}$ is a way to simulate diagonal disorder in a quantum mechanical one-dimensional tight-binding Hamiltonian, where the coupling $\eta$ between nearest neighbors is a constant~\cite{SorathiaIzrailevZelevinskyCelardo}. In this case the Anderson localization phenomenon also occurs.

We shall now compute and measure the localization length $\xi$ for a disordered rod.  This quantity can be defined in at least three ways~\cite{KramerMacKinnon}: using the exponential decay of the transmission coefficient; by the Lyapunov exponent of the transfer matrix, or by the exponential decay of the wave amplitude envelope. Because the wave functions are in general not accessible experimentally, the first two definitions are normally used in the literature. However, we do have access to the wave functions, so we shall use the last definition.

To obtain the experimental localization length $\xi$ from the wave amplitudes, as those shown in Fig.~\ref{Fig.WaveAmplitudes}, we consider a frequency interval and measure all the amplitudes with an eigenfrequency within this interval. An exponential is fitted by a least-squares procedure to the amplitude envelope
\begin{equation}
 \psi_\mathrm{env} = A \exp \left(- \frac{|x-x^*|}{\xi} \right)
\end{equation}
at $x>x^*$ where $A$ is a constant. Here $x^*$ and $\xi$ are the position of the maximum and the localization length of the wave amplitude, respectively. When the wave amplitude presents more than one maxima, only the highest one is used in the least-squares procedure. In this way the localization length was obtained as a function of the normal-mode frequency. To calculate the corresponding numerical wave amplitudes the Poincar\'e map method~\cite{AvilaMendez-Sanchez} was used and $\xi$ was obtained in a similar fashion as before. It should be remarked that, as has been the case in all numerical calculations we have previously compared with experimental results, an effective value of the coupling parameter $\eta$ is necessary (See Ref.~\cite{Moralesetal} where this is justified). The calculations and the experiment coincide extremely well with each other using only this adjustable parameter.

\begin{figure}
\includegraphics[width=\columnwidth]{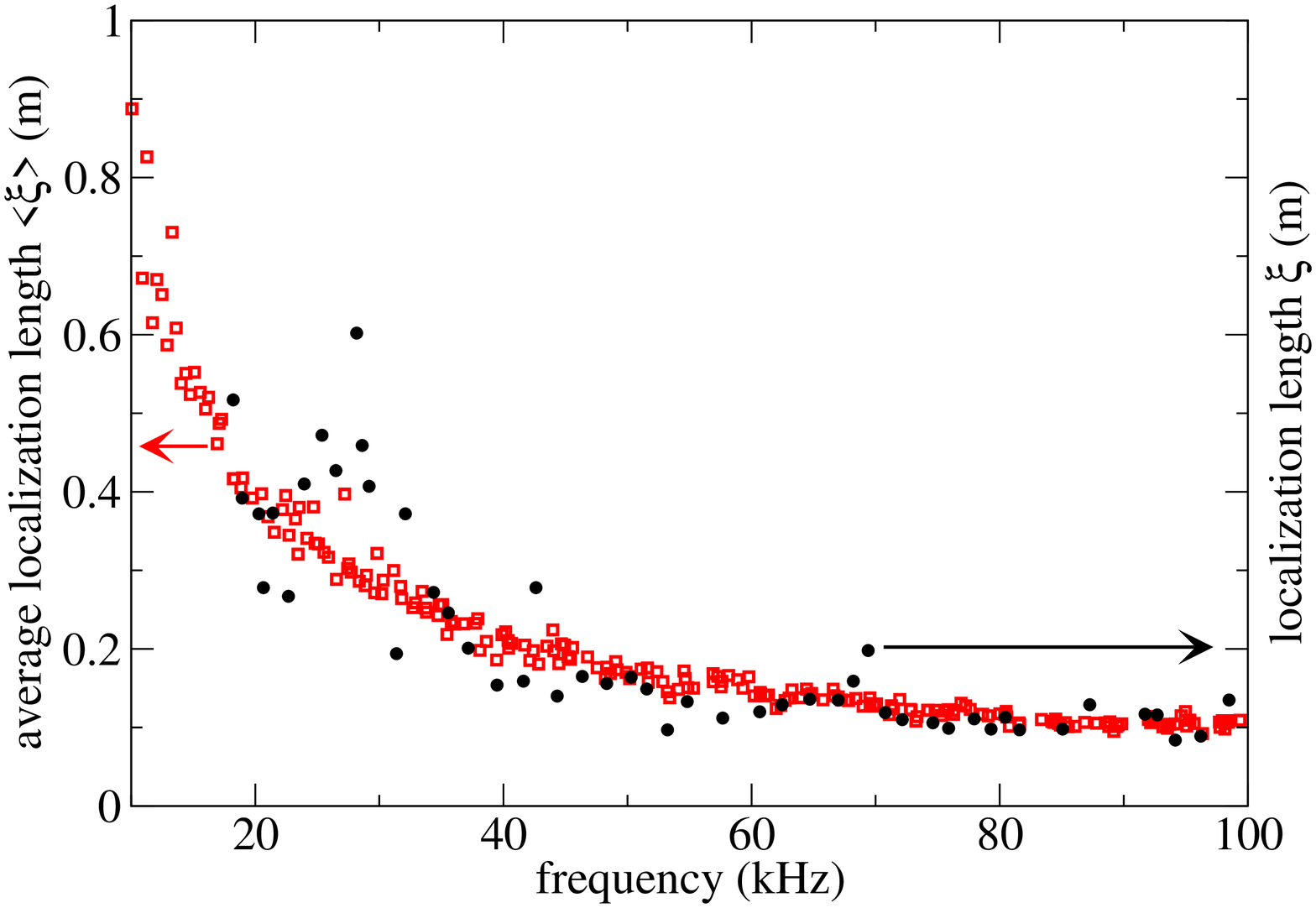}
\caption{(Color online) Average of the localization length (squares) as a function of the normal-mode frequency. The frequency average was done in windows of $20$kHz. The experimental localization length for a single rod is given by the dots. It should be noted that $\xi$ decreases as the frequency $f$ grows. }
\label{Fig.LocalizationLength}
\end{figure}

The localization length $\xi$ as a function of frequency is given in Fig.~\ref{Fig.LocalizationLength}. One should mention the overall agreement between theory (squares) and experiment (dots). The numerical values are obtained from an ensemble of 5000 families $\{d_i\}$ but the experimental values were measured only for 50 eigenfunctions of a single rod. It is to be noted that $\xi$ decreases with frequency.

To obtain the wave amplitudes, such as those shown in Fig.~\ref{Fig.WaveAmplitudes}, the spectrum of the disordered rod must first be obtained. This is the case both numerically and in the laboratory. We are then provided with an extra bonus: the statistical properties of the elastic spectra which render themselves to studies like those analyzed in random matrix physics and in quantum chaos~\cite{Brodyetal,Guhretal}. In what follows we shall provide the nearest-neighbor spacing distribution $p(s_i)$, where $s_i = \frac{f_{i + 1}- f_i}{\langle f_{i+1}-f_i\rangle}$ is the normalized spacing, and show how the distribution varies as a function of frequency.

\begin{figure}
\includegraphics[width=\columnwidth]{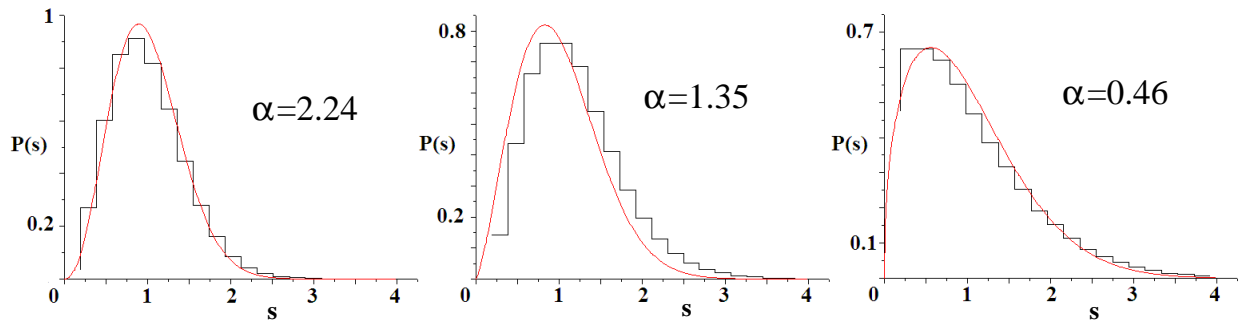}
\caption{(Color online) The nearest-neighbor spacing distribution $p(s)$, obtained for an ensemble of 5000 rods and average over a window frequency interval of 20~kHz, changes with frequency. From left to right the center of the interval is 28~kHz, 38~kHz and 78~kHz, respectively.} 
\label{Fig.NumericalNNSDs}
\end{figure}

\begin{figure}
\includegraphics[width=\columnwidth]{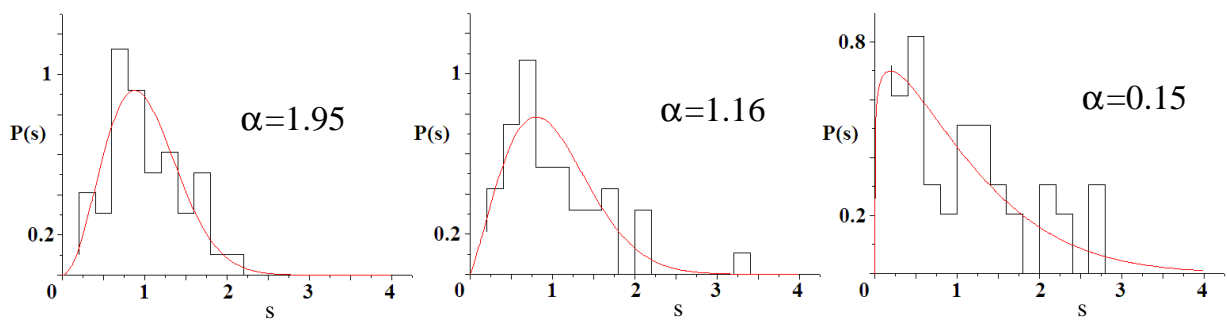}
\caption{(Color online) The nearest-neighbor spacing distribution $p(s)$ obtained experimentally, with an average over a window frequency interval of 20~kHz, for a typical rod varies with the frequency. From left to right the center of the interval is 20~kHz, 24~kHz and 48~kHz, respectively.}
\label{Fig.ExperimentalNNSDs}
\end{figure}

We first consider an ensemble of 5000 disordered rods from a numerical point of view.  The richness of the elastic 1D system shows itself off when trying to fit the nearest-neighbor spacing distribution along the spectrum. We see that a more general function is needed than those calculated for the Gaussian orthogonal (GOE), the Gaussian unitary (GUE) and the Gaussian symplectic (GSE) random matrix ensembles. As a matter of fact, to obtain the level repulsion parameter $\alpha$, the distribution
\begin{equation}
 p_{\alpha}(s) = A s^{\alpha}(1+B \alpha s)^{f(\alpha)}
\exp{\left[-\frac{\pi^2}{16}\alpha s^2-\frac{\pi}{2}\left(1-\frac{\alpha}{2}\right)s\right]},
\end{equation}
where 
\begin{equation}
 f(\alpha)=\frac{2^\alpha\left(1-\frac{\alpha}{2}\right)}{\alpha}-0.16874,
\end{equation}
and $A$, $B$ are constants such that the distribution and the average of the spacing are normalized~\cite{Izrailev}, was fitted to the numerical data by means of a least-squares procedure. Examples of numerical $p_{\alpha}(s)$ for different regions of the frequency spectrum are shown in Fig.~\ref{Fig.NumericalNNSDs}. The repulsion parameter $\alpha$ varies with the frequency. This is observed also in the experimental values of $\alpha$ shown in Fig.~\ref{Fig.ExperimentalNNSDs}.

Given these results one could also consider the relationship of $\xi$ with $\alpha$. This has been done when dealing with a disordered quasi-1D wire in mesoscopic physics. The Dorokhov-Mello-Pereyra-Kumar (DMPK) equation was derived~\cite{DMPK} as well as a relationship between the localization length, the elastic mean free path $\ell_e$, and the Dyson parameter $\beta=1,2,4$, which characterize the Gaussian orthogonal, unitary and sympletic ensembles, respectively. The Dyson parameter $\beta$ is also associated to the nearest-neighbor spacing distribution of the quantum energy spectrum. Indeed $\alpha=\beta$ for $\beta=1,2,4$. It was shown~\cite{StoneMelloMuttalibPichard} that a linear relationship between $\xi$ and $\beta$ exists
\begin{equation}
 \xi=\left[( N_\mathrm{c}-1)\beta +2\right] \ell_e
 \label{Eq.LocalizationLength}
\end{equation}
where $N_\mathrm{c}$ is the total number of channels (for a review see~\cite{Beenakker}). Eq.~(\ref{Eq.LocalizationLength}) was verified for $\beta=1,2,4$ in magnetoconductance experiments~\cite{PichardSanquerSlevinDebray}.

\begin{figure}
\includegraphics[width=\columnwidth]{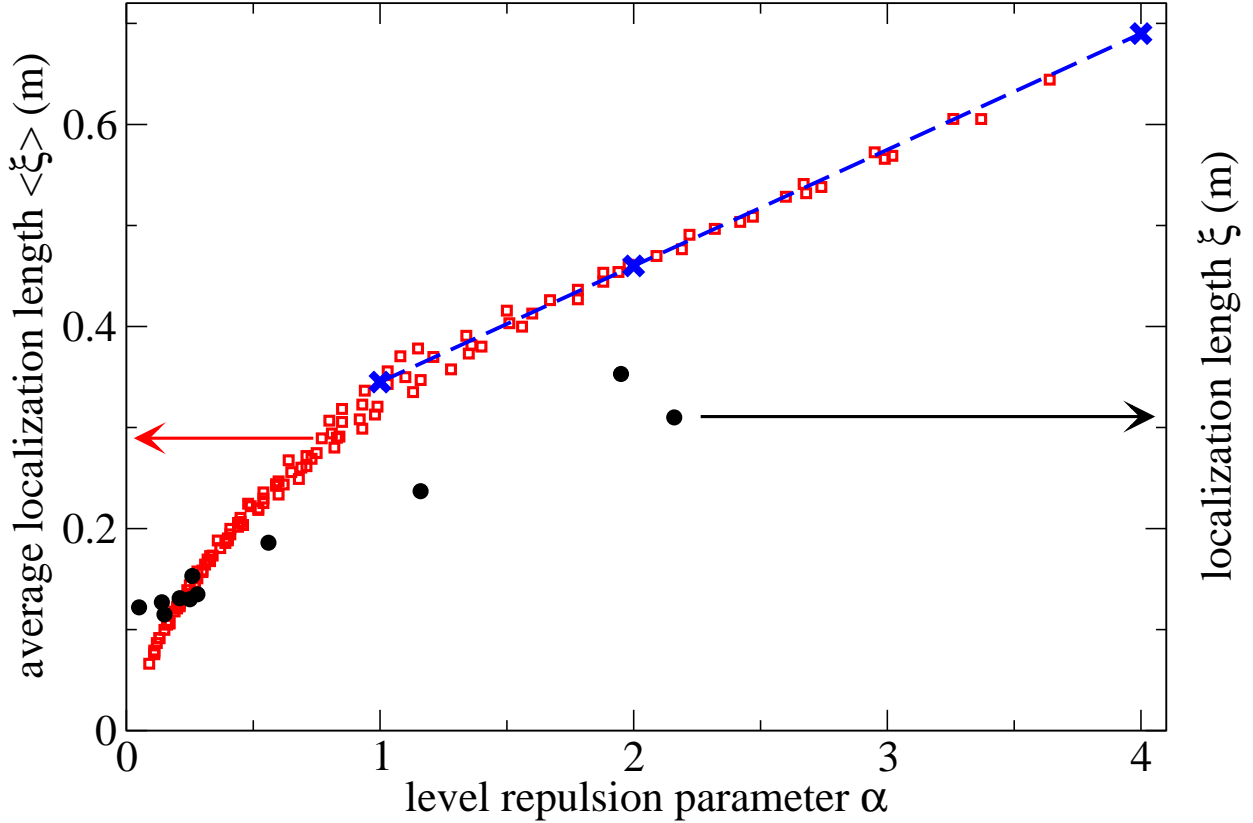}
\caption{(Color online) Average localization length $\xi$ (squares) as a function of the repulsion parameter $\alpha$. The values $\alpha =1,2,4$ (crosses) correspond to the random matrix theory predictions, with $N_\mathrm{c}=2$, $\ell_e=0.115$~m, for which $\xi$ is a linear function of $\alpha$ (dashed line). For $\alpha < 1$ this dependence is no longer linear. The dots correspond to the experimental values obtained for a single rod.}
\label{Fig.LocalizationLengthVsAlpha}
\end{figure}

The plot $\xi$ {\em vs} $\alpha$ for the elastic rods is given in Fig.~\ref{Fig.LocalizationLengthVsAlpha}. The points were obtained for the ensemble of rods. For $\alpha = 1,2,4$ this repulsion parameter coincides with $\beta$; one can see from the crosses in this figure that indeed $\xi$ is a linear function of $\alpha$. But this is not true at low values of $\alpha$, where a higher order polynomial fits the theoretical values better. The experimental results for our rod also show this, which is a new effect, not obtained with the usual random matrix ensembles. 

A qualitative explanation for the decrease of $\xi$ when $\alpha\rightarrow 0$  is again provided by the independent rod model. When $\eta\rightarrow 0$, the frequency spectrum, defined by the independent random sequence $\{d_i\}$, is also random. As is well known in random matrix theory the nearest-neighbor spacing resembles then a Poisson distribution, that is, the repulsion parameter tends to zero. On the other hand, $\xi$ also diminishes in this case, since the wave amplitudes are localized in a single small rod. We therefore understand, from a qualitative point of view, the structure of the curve shown in Fig.~\ref{Fig.LocalizationLengthVsAlpha}.

To conclude, in this Letter we measured the exponential localization of elastic wave amplitudes in disordered rods as well as the spectrum and calculated the level repulsion parameter $\alpha$. It was found that the localization length presents two regimes. For $\alpha>1$ the localization length grows linearly as a function of the level repulsion parameter. This agrees with the predictions of the DMPK equation for the Dyson random matrix ensembles. However, there is a region, $\alpha<1$, in which the localization length is not a linear function of $\alpha$. 

This work is supported by DGAPA-UNAM under project IN111311. RAMS was supported by CONACYT under project 79613.

\end{document}